\documentclass[journal=jacsat,manuscript=article]{achemso}

\usepackage{chemformula} % Formula subscripts using \ch{}
\usepackage[T1]{fontenc} % Use modern font encodings

\author{Hendrik Meer}
\email{meer@uni-mainz.de}
\affiliation{Institute of Physics, Johannes Gutenberg-University Mainz, 55099 Mainz, Germany}

\author{Stephan Wust}
\affiliation{Department of Physics and Research Center OPTIMAS, Technische Universit{\"a}t Kaiserslautern, 67663 Kaiserslautern, Germany}

\author{Christin Schmitt}
\affiliation{Institute of Physics, Johannes Gutenberg-University Mainz, 55099 Mainz, Germany}

\author{Paul Herrgen}
\affiliation{Department of Physics and Research Center OPTIMAS, Technische Universit{\"a}t Kaiserslautern, 67663 Kaiserslautern, Germany}
    
\author{Felix Fuhrmann}
\affiliation{Institute of Physics, Johannes Gutenberg-University Mainz, 55099 Mainz, Germany}
    
\author{Steffen Hirtle}
\affiliation{Department of Physics and Research Center OPTIMAS, Technische Universit{\"a}t Kaiserslautern, 67663 Kaiserslautern, Germany}

\author{Beatrice Bednarz}
\affiliation{Institute of Physics, Johannes Gutenberg-University Mainz, 55099 Mainz, Germany}

\author{Adithya Rajan}
\affiliation{Institute of Physics, Johannes Gutenberg-University Mainz, 55099 Mainz, Germany}
    
\author{Rafael Ramos}
\affiliation{WPI-Advanced Institute for Materials Research, Tohoku University, Sendai 980-8577, Japan}
\altaffiliation{Centro de Investigación en Química Biolóxica e Materiais Moleculares (CIQUS), Departamento de Química-Física, Universidade de Santiago de Compostela, Santiago de Compostela 15782, Spain}

\author{Miguel Angel Niño}
\affiliation{ALBA Synchrotron Light Facility, Carrer de la Llum 2-26, Cerdanyola del Vallés, 08290 Barcelona, Spain}
\author{Michael Foerster}
\affiliation{ALBA Synchrotron Light Facility, Carrer de la Llum 2-26, Cerdanyola del Vallés, 08290 Barcelona, Spain}
\author{Florian Kronast}
\affiliation{Helmholtz-Zentrum Berlin für Materialien und Energie, Albert-Einstein-Strasse 15, 12489 Berlin, Germany}
\author{Armin Kleibert}
\affiliation{Swiss Light Source, Paul Scherrer Institut, 5232 Villigen PSI, Switzerland}
\author{Baerbel Rethfeld}
\affiliation{Department of Physics and Research Center OPTIMAS, Technische Universit{\"a}t Kaiserslautern, 67663 Kaiserslautern, Germany}
\author{Eiji Saitoh}
\affiliation{WPI-Advanced Institute for Materials Research, Tohoku University, Sendai 980-8577, Japan}
\alsoaffiliation{Department of Applied Physics, The University of Tokyo, Tokyo 113-8656, Japan}
\alsoaffiliation{Center for Spintronics Research Network, Tohoku University, Sendai 980-8577, Japan}
\alsoaffiliation{Advanced Science Research Center, Japan Atomic Energy Agency, Tokai 319-1195, Japan}
\author{Benjamin Stadtmüller}
\affiliation{Institute of Physics, Johannes Gutenberg-University Mainz, 55099 Mainz, Germany}
\alsoaffiliation{Department of Physics and Research Center OPTIMAS, Technische Universit{\"a}t Kaiserslautern, 67663 Kaiserslautern, Germany}
\author{Martin Aeschlimann}
\affiliation{Department of Physics and Research Center OPTIMAS, Technische Universit{\"a}t Kaiserslautern, 67663 Kaiserslautern, Germany}
\author{Mathias Kl\"aui}
\email{Klaeui@Uni-Mainz.de}
\affiliation{Institute of Physics, Johannes Gutenberg-University Mainz, 55099 Mainz, Germany}
\title
  {Laser-induced Creation of antiferromagnetic 180-degree domains in NiO/Pt bilayers}
\usepackage{atbegshi}% http://ctan.org/pkg/atbegshi
\AtBeginDocument{\AtBeginShipoutNext{\AtBeginShipoutDiscard}}

\begin{document}
\begin{abstract}

We demonstrate how the antiferromagnetic order in heterostructures of NiO/Pt thin films can be modified by optical pulses. We irradiate our samples with laser light and identify an optically induced creation of antiferromagnetic domains by imaging the created domain structure utilizing the X-ray magnetic linear dichroism effect. We study the effect of different laser polarizations on the domain formation and identify a polarization-independent creation of 180$^\circ$ domain walls and domains with 180$^\circ$ different Nèel vector orientation. By varying the irradiation parameters, we determine the switching mechanism to be thermally induced and demonstrate the reversibility. We thus demonstrate experimentally the possibility to optically create antiferromagnetic domains, an important step towards future functionalization of all optical switching mechanisms in antiferromagnets.
\end{abstract}


\section{Introduction}
To overcome the limitations of ferromagnetic systems (FMs) it is a key step to transition to antiferromagnetic systems (AFMs) in future spintronic devices. Due to their net magnetic moment FMs are limited in their bit packing density and stability against external fields compared to antiferromagnets \cite{Baltz2018}. Another key advantage of antiferromagnets is their potential for ultrafast applications due to their inherent dynamics with resonant frequencies in the THz range \cite{Kampfrath2011}. Recently, the electrical switching of antiferromagnets has been intensely investigated \cite{Wadley2016, Bodnar2018, Moriyama2018}. However, to achieve switching on an ultrashort timescale, we must transition from electrical to optical modification of the antiferromagnetic order.

In ferri- and ferromagnetic material systems, fs-laser-induced all-optical switching (AOS) has been intensively studied \cite{Stanciu2007,ElHadri2017,Alebrand2012,Kimel2019}. Thermally induced switching has been observed in ferrimagnetic GdFeCo alloys \cite{Stanciu2007,Radu2011} and all-optical helicity-dependent switching (AO-HDS) based on the inverse Faraday effect has been observed in a wide range of ferri- and ferromagnetic materials \cite{Mangin2014}. For antiferromagnetic materials, studies on the all-optical switching have focused on the excitation of magnon modes \cite{Kimel2004,Tzschaschel2017,Kanda2011,Nemec2018}. There are experimental reports of large-scale optical switching of antiferromagnetic order in the tilted antiferromagnet TbMnO$_3$ \cite{Manz2016}. However, the underlying mechanism relies on the electric polarization and cannot be easily transferred to other AFM systems. Theoretical studies predict possible optically induced switching in antiferromagnetic NiO \cite{Lefkidis2007,Lefkidis2009}, NiO/FM bilayers \cite{Chirac2020a} and other collinear antiferromagnets \cite{Dannegger2021a}. Recently, first experimental evidence for light-induced manipulation of antiferromagnetic domains in NiO crystals has been reported \cite{Stremoukhov2022}. NiO is a prototypical collinear insulating antiferromagnetic system exhibiting promising features for future potential spintronic devices: current-induced switching of the antiferromagnetic order \cite{Moriyama2018, Meer2021}, electrical readout \cite{Hoogeboom2017}, antiferromagnetic shape anisotropy \cite{Meer2022}, and ultrafast spin dynamics in the THz range \cite{Kampfrath2011,Higuchi2011,Rongione2022}. The magnetic order of NiO thin films can be temporarily modulated by irradiation with ultrafast laser pulses \cite{Wust2022}, and several studies have reported helicity-dependent excitation of coherent magnons in NiO \cite{Satoh2010a, Tzschaschel2017}. NiO exhibits a strong magnetoelastic coupling \cite{Aytan2017} and thus the above-mentioned optical manipulation of antiferromagnetic domains \cite{Stremoukhov2022} has been attributed to a particular phononic mechanism \cite{Stupakiewicz2021}. While several mechanisms have theoretically proposed optically induced switching of NiO, there are no experimental reports on the light-induced domain switching of the antiferromagnetic order in NiO thin films. 

Here, we investigate the domain structure of NiO/Pt bilayers using X-ray photo-emission electron microscopy (XPEEM) with magnetic linear dichroism as the contrast mechanism, that were irradiated by circularly and linearly polarized laser-light. We observe optically induced changes of the domain structure. In contrast to often considered switching between different Néel vector axes, we observe the creation of 180$^\circ$ domains and domain walls, independent of the laser polarization. Variation of the irradiation parameters allows us to identify a thermal origin of the optically induced antiferromagnetic order. We demonstrate the possibility to optically create antiferromagnetic domains in a prototypical antiferromagnetic system.

\section{Results} 
We prepared 10$\,$nm thick NiO(001) thin films that are epitaxially grown on MgO(001) substrates by reactive magnetron sputtering. The films are additionally capped with a 2$\,$nm thick platinum layer to allow for imaging with X-ray magnetic linear dichroism photoemission electron microscopy (XMLD-PEEM). We have previously characterized and investigated similarly grown NiO(001) thin films and observed that the strain from the substrate mismatch leads to preferential out-of-plane alignment of the Néel vector in these thin films, stabilizing only one type of Spin-domains (S-domain) \cite{Baldrati2019,Alders1998, Altieri2003, Schmitt2020b}. Therefore, only four different Twin-domains (T-domains) are present in our films, each accompanied by a strong rhombohedral distortion. Similar to our previous studies on 10$\,$nm NiO thin films \cite{Schmitt2020b}, we observe a domain structure in our field of view (FOV) which predominantly consists of one T-domain.
To study the effect of laser-irradiation on NiO/Pt bilayers, we use an ultrafast amplified laser system with a central wavelength of 800$\,$nm and with a pulse repetition rate of 1$\,$kHz. We irradiated our samples with pulse trains of different pulse duration, pulse fluences, illumination time, and polarization. 
We imaged the antiferromagnetic domain structure of the laser irradiated regions using energy dependent XMLD-PEEM at the double peak of the Ni L$_2$ edge \cite{Stohr1999}.
Fig. 1 shows the antiferromagnetic domain structure of a region irradiated with circularly right-polarized laser light.
\begin{figure}[h]
\includegraphics[width=0.45\textwidth]{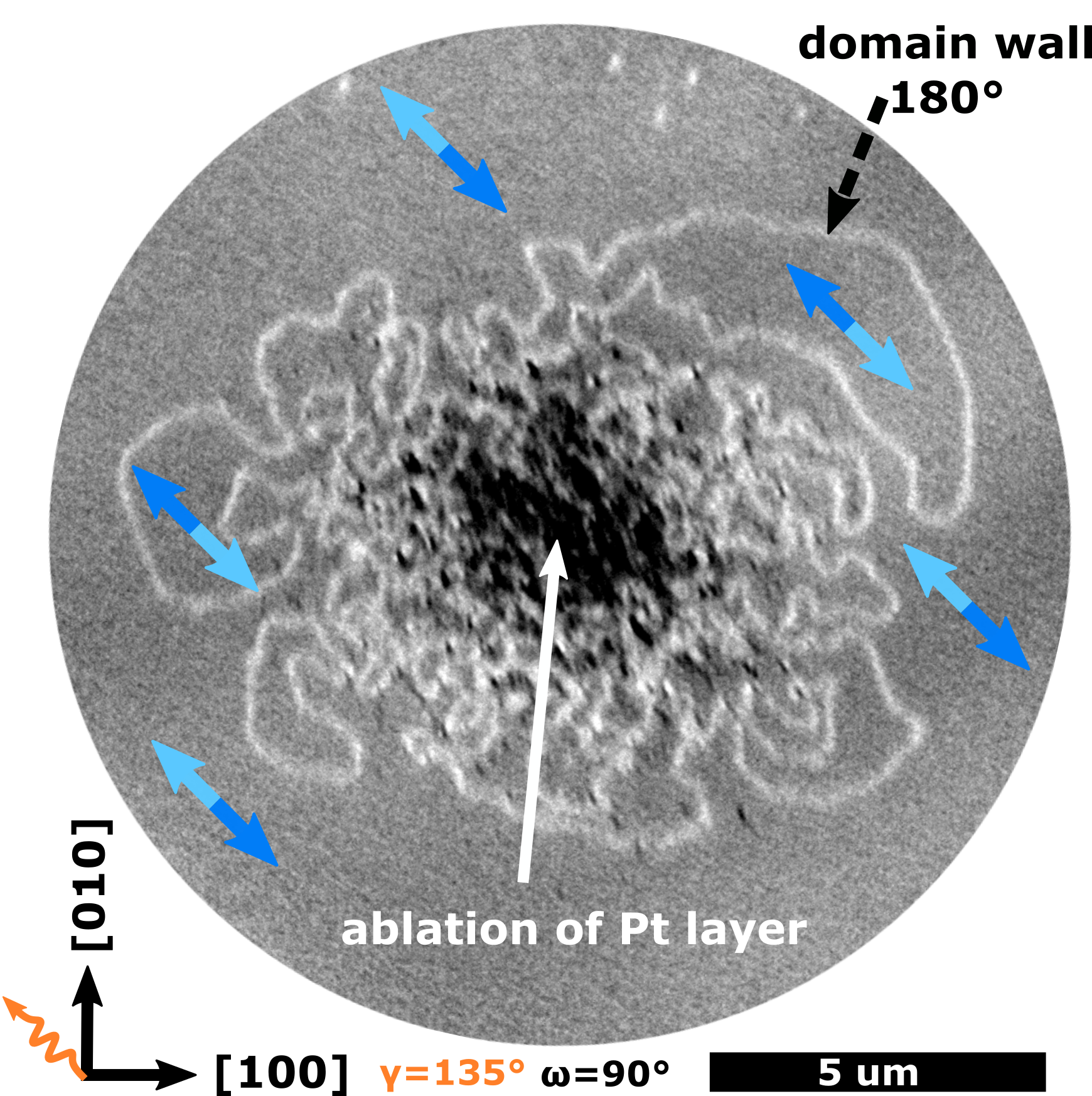}
\caption{\label{fig:1} XMLD-PEEM image of the AFM domain structure, of an area that was illuminated for 2\,s with circular right polarized laser light with a pulse length of 6.7$\,$ps and a fluence of 12$\pm$2$\,$mJ$/$cm$^2$. The incoming X-ray is at an angle of $135^\circ$ with respect to the [100] axis and the X-ray polarization is vertical to the sample plane with a $16^\circ$ incidence angle. The blue arrows indicate the direction of the in-plane projection of the Néel vector, the different shades correspond to the different sublattices.}
\end{figure}
Several narrow domain walls (bright lines) can be observed around the central laser spot. The XMLD contrast depends on the orientation of the incoming electric field and the orientation of the Néel vector. We varied the polarization of the incoming X-ray ($\omega=90^\circ,112.5^\circ,135^\circ,157.5^\circ,0^\circ$) and the azimuthal angle ($\gamma=90^\circ,120^\circ,135^\circ,150^\circ,165^\circ,180^\circ$) of the sample with respect to the incoming beam. For all combinations of $\omega$ and $\gamma$ we observe no difference in contrast between the domains inside and outside the domain walls. The absence of contrast changes indicates that the projection of the Néel vector in these domains onto the X-ray polarization is identical for all angles of $\omega$ and $\gamma$ and points along the same directions \cite{Czekaj2007a}.
However, for different combinations of $\omega$ and $\gamma$ we could observe variation and inversion of the contrast between the domain wall and the surrounding domains (see Supporting Information). The contrast of the domain wall appears uniform, as previously reported in our recent study on domain walls between T-domains in NiO thin films \cite{Schmitt2022}. As the projection of the Néel vector between both the domains is identical, the created domain wall can be identified as a 180$^\circ$ domain wall between two 180$^\circ$ domains. The orientation of the Néel vector is identical on both sides of the wall, but the spins in the antiferromagnetically coupled sublattices are interchanged, as indicated by the differently shaded blue arrows in Fig. 1. Thus, the domains inside and outside of the domain walls are in the same T-domain, are accompanied by the same distortion, and their S-domains have the same out of plane components. But their sublattices are interchanged, making them 180$^\circ$ differently oriented domains. By irradiation with a laser, we are able to observe the creation of 180$^\circ$ domain walls (bright lines in Fig.1), which indicate the creation of domains with 180$^\circ$ different orientation of the Néel vector.

The creation of 180$^\circ$ domains is independent of the polarization of the laser and can also be achieved with linearly polarized laser light (see Fig. 2a).

\begin{figure}[h]
\includegraphics[width=0.45\textwidth]{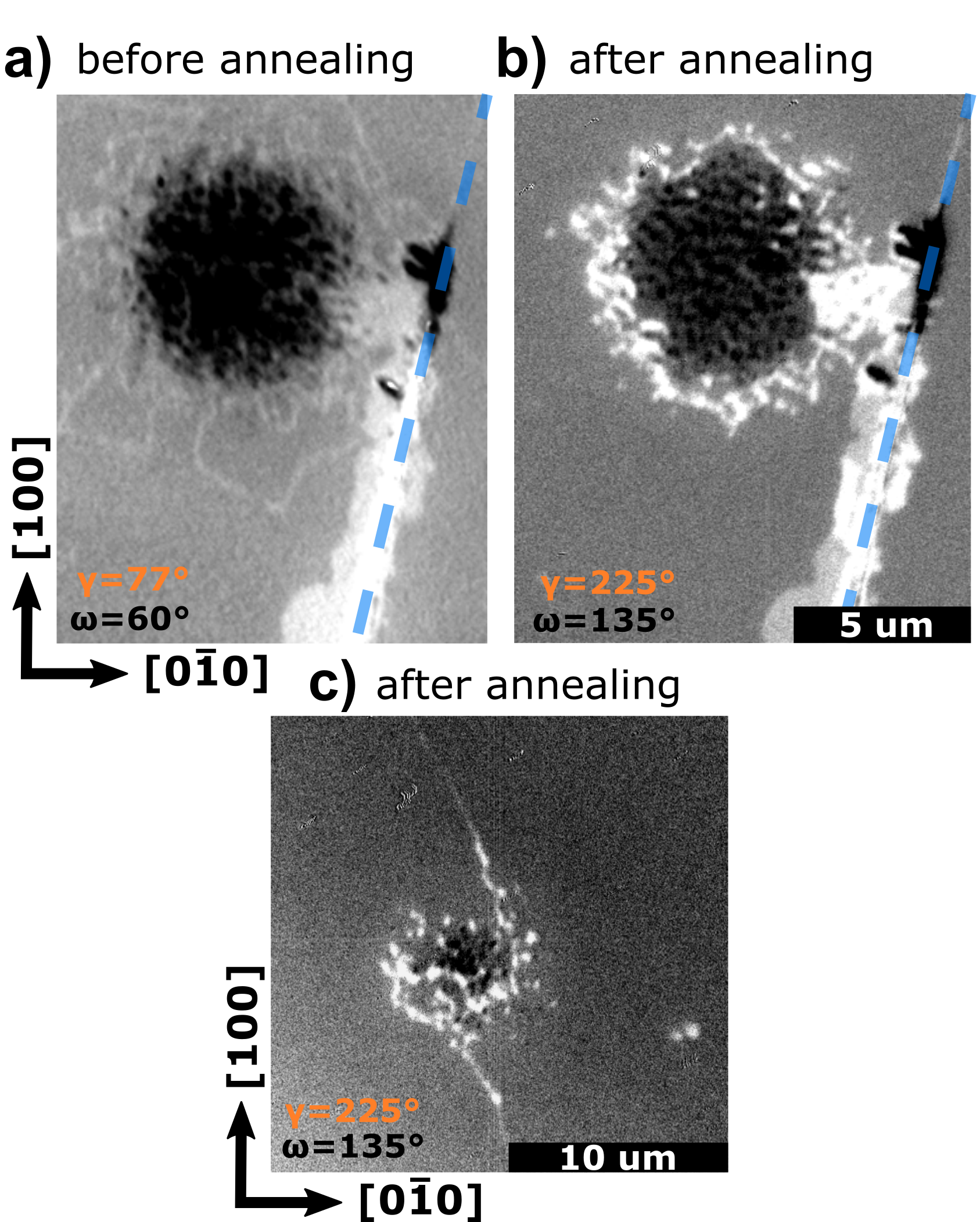}
\caption{\label{fig:2} (a) XMLD-PEEM image of an area that was illuminated for 2\,s with linearly polarized laser light with a pulse length of 6.7$\,$ps, 2000 pulses and a fluence of 16$\pm$2$\,$mJ$/$cm$^2$. (b) Domain structure of the same laser-irradiated area after annealing above the Néel temperature. The blue line indicates a structural defect at which domains are stabilized. (c) XMLD-PEEM image of the area depicted in Fig. 1. after annealing.}
\end{figure}

To verify the antiferromagnetic nature of the laser-induced domains, we annealed our film in vacuum for 10 minutes without magnetic field above Néel temperature, at 550 K. This has changed the domain structure and we can observe a disappearance of the 180$^\circ$ domain walls in Fig. 2b. In Fig. 2c we imaged the area from Fig. 1 after annealing and we can also observe that most 180$^\circ$ domain walls have disappeared. We can observe one 180$^\circ$ domain wall going across the laser irradiated area, which changes contrast depending on beam polarization (see Supplementary Information).

As discussed above, the laser-induced 180$^\circ$ domains in Fig. 2b have disappeared. However, the domains near the structural defect were not significantly altered by the heating. 
In NiO, strain or changes in the surface anisotropy that are introduced by patterning or defects can lead to the preferential stabilization of domains \cite{Meer2022}. The optically induced 180$^\circ$ domains are not stabilized and can be manipulated. Thus, the laser-induced domains are not generated by irreversible ablation-induced defects, which introduce strain or change the surface anisotropy, but originate from rapid heating and cooling of the antiferromagnetic NiO system. The laser-induced domain creation is dominated by thermal processes, as we can observe polarization independent creation, which occurs close to the ablation threshold. Only the accumulated heating and ablation threshold differ slightly between different laser polarizations \cite{Venkatakrishnan2002a}. For linear polarized light under the same conditions as in Fig. 2a, but with a lower fluence of 12$\pm$2$\,$mJ$/$cm$^2$, we could not observe a visible ablation or the creation of 180$^\circ$ domain walls, indicating that the accumulated heat was not sufficient to allow a reorientation of the spins. 
We can estimate the laser-induced temperature increase during irradiation by considering the melting temperature of Pt as a lower boundary for the temperature increase near the ablated area. The temperature of the NiO near ablated areas can be estimated to be around 520$\,$K, which is slightly below the bulk NiO Néel temperature, but above the reduced Néel temperature in our thin film ($T_\text{N,thin}=400\,$K-460$\,$K, see Supplementary Information). Thus, the laser-induced heating near the ablation threshold can be assumed to be sufficient to allow a reconfiguration of the domain structure.
In the case that the laser-induced domain creation is of thermal origin, it can be achieved without structurally affecting the Pt layer.

\begin{figure}[h]
\includegraphics[width=0.45\textwidth]{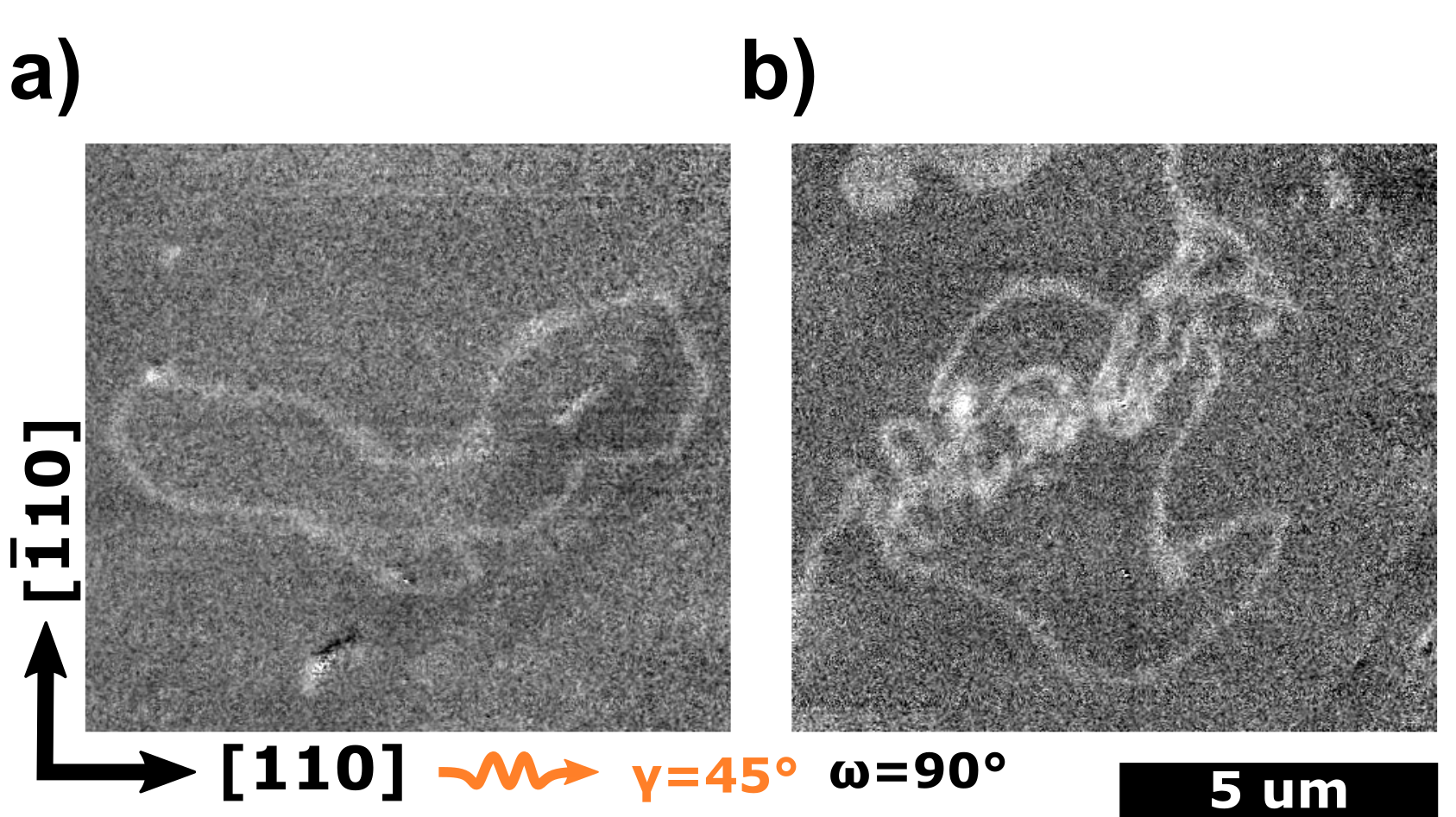}
\caption{\label{fig:3} XMLD-PEEM image of two area that were illuminated for 0.5\,s with linearly polarized laser light with a pulse length of 45$\,$fs and a fluence of 8$\pm$1$\,$mJ$/$cm$^2$ (a) or 7.2$\pm$0.9$\,$mJ$/$cm$^2$ (b).}
\end{figure}
To create domains without damaging the Pt layer, we next investigate different laser pulse parameters. We patterned grids of markers onto the sample and irradiated the sample in the center of the grids. This allows us to identify the laser irradiated region in the XPEEM even if there is no visible damage to the Pt layer. We could observe the creation of domains without visible damage to the Pt layer using 45$\,$fs pulses and irradiation with 500 pulses over 0.5 seconds. The characteristic formation of 180$^\circ$ domain walls could be observed, as shown in Figs. 3a and 3b, for irradiation with linearly polarized light. The threshold fluence for the formation of 180$^\circ$ domain walls under these irradiation conditions is found to be 7.2$\pm$0.9$\,$mJ$/$cm$^2$. Thus, we can optically create antiferromagnetic 180$^\circ$ domains and domainwalls without damaging the NiO or the platinum capping layer. 

\section{Discussion}

We have observed the optically induced formation of $180^\circ$ domain walls in NiO, indicating the creation of $180^\circ$ domains. We observe polarization-independent creation of the antiferromagnetic domains. The underlying mechanism is not based on a polarization dependent effect but on heat-induced domain formation, since the nucleation of the 180° domains is independent of the polarization of the exciting light. 
Irradiation with the laser light heats up the NiO and its spin system. The lattice system around the heated area is still strained, exerting strain on the system that is heated up. During cooling of the irradiated area, this strain leads to preferential alignment of the spins into the previous twin domain after irradiation. The heated disordered spins settle into one of the two energetically degenerate states, with spins in the different sublattices pointing in opposite directions. Thus, 180$^\circ$ domains and 180$^\circ$ domain walls are formed. The situation is different when the whole sample is annealed, as shown in Fig. 2b. As the creation of 180$^\circ$ domain walls requires additional energy, it is energetically unfavourable. In the slow cooling process of the annealing, (3.0±0.2$\,$K/min, see Supplementary information), the domain wall formation is avoided and their size is minimized. These results are consistent with observations on bulk NiO: it was observed that a rapid cooling through the Néel temperature (opening the furnace door) can lead to a multidomain state, while annealing with slow cooling (5$\,$K/min) leads to crystals with only few domains \cite{Armstrong2009}.
The size of the optically induced domains is not necessarily linear with fluence, as can be seen in Fig. 3a and Fig. 3b, but depends strongly on the local anisotropy landscape of the irradiated area. However, by tuning the timescale of the laser heating pulse, one can potentially switch reversibly between different magnetic states.

Previous experiments based on current-induced switching have allowed for the controlled switching between different T-domains \cite{Schmitt2020b}. The laser-induced domain creation offers the exciting possibility to additionally create and study 180$^\circ$ domain walls, which have recently been reported to be rather narrow and even atomically sharp in antiferromagnets \cite{Krizek2022}.
Further, domain walls themselves can play a crucial role in the magnon coupling of the NiO modes \cite{Bossini2021a,Gomonay2021a}. By controlled irradiation of a device, one could artificially introduce additional domain walls into the system, decreasing the size of the individual domains. In this way, laser-induced domain creation can be used to locally lower the potential switching energy for a device or to manipulate magnon transport inside a device.

While all-optical creation of 180$^\circ$ domains itself offers exciting opportunities, we believe that the demonstration of optically induced creation of large domains in NiO is an important step for the investigation of AOS in AFMs. The pulse duration is found to be a crucial parameter in this process, since the underlying mechanism is based on the rapid heating and the subsequent cooling of the system. Therefore, there is potential to push the writing speed to faster regimes by optimizing this optical parameter, while simultaneously tailoring the material parameters of the sample. Additionally, the variation of irradiation parameters could provide a tool for the realization of reversible optical switching. Different heat cycles could be used to switch between multi- and mono-domain states, similar to phase-change memory \cite{Pohm1970}.

It should be emphasized that our observation of a heat-induced optical domain formation does not imply the absence of other optical switching mechanisms in antiferromagnets or NiO. In the experiment presented here, we have observed 180$^\circ$ domain creation inside one T-domain, due to the preferential stabilization of only one S-domain, in our NiO (001) thin films \cite{Schmitt2020b}. In other systems with other domain configurations the optically induced domain formation might not be as restrained. To investigate the all optical switching in antiferromagnets one has to therefore carefully choose the material platform, a promising candidate might be NiO(111).
In NiO (111) are three different S-domains and a total of six different magnetic states per T-domain present. Thus, NiO (111) could be the ideal platform to explore the theoretical proposed optical switching mechanisms, as the spins could switch between multiple S-domains \cite{Chirac2020a}. 
We believe that the mechanism found here can also be achieved in other antiferromagnetic systems with strong magnetoelastic coupling, such as CoO or Hematite. The optically induced creation of 180$^\circ$ domains and domain walls thus provides an additional handle to manipulate antiferromagnetic domain structures.

\begin{acknowledgement}

The authors thank T. Reimer and L. Schnitzspan for skillful technical assistance and H. Gomonay for her comments. Experiments were performed at the CIRCE beamline at ALBA Synchrotron with the collaboration of ALBA staff. We thank HZB for the allocation of synchrotron radiation beamtime, we thankfully acknowledge the financial support by HZB. We acknowledge the Paul Scherrer Institute, Villigen, Switzerland for the beamtime allocation under proposal 20211822 at the SIM beamline of the SLS. B.B. and A. R. acknowledge funding from the European Union’s Framework Programme for Research and Innovation Horizon 2020 (2014-2020) under the Marie Skłodowska-Curie Grant Agreement No. 860060 (ITN MagnEFi). M.K. acknowledges support from the Graduate School of Excellence Materials Science in Mainz (MAINZ) DFG 266, the DAAD (Spintronics network, Project No. 57334897 and Insulator Spin-Orbitronics, Project No. 57524834), and all groups from Mainz and Kaiserslautern acknowledge that this work was funded by the Deutsche Forschungsgemeinschaft (DFG, German Research Foundation), TRR 173-268565370 (Project Nos. A01, A08, B02 and B03) and KAUST (OSR-2019-CRG8-4048). B.S. further acknowledges funding by the Dynamics and Topology Research Center (TopDyn) funded by the State of Rhineland Palatinate. R.R. also acknowledges support from the European Commission through the Project 734187-SPICOLOST (H2020-MSCA-RISE-2016), the European Union's Horizon 2020 research and innovation program through the MSCA Grant Agreement SPEC No. 894006, Grant RYC 2019-026915-I funded by the MCIN/AEI/10.13039/501100011033 and by "ESF investing in your future", the Xunta de Galiciaa (ED431F 2022/04, ED431B 2021/013, Centro Singular de Investigación de Galicia Accreditation 2019-2022, ED431G 2019/03) and the European Union (European Regional Development Fund - ERDF). M.K. acknowledges financial support from the Horizon 2020 Framework Programme of the European Commission under FET-Open Grant Agreement No. 863155 (s-Nebula). This work was also supported by ERATO “Spin Quantum Rectification Project” (Grant No. JPMJER1402) and the Grant-in-Aid for Scientific Research on Innovative Area, “Nano Spin Conversion Science” (Grant No. JP26103005), Grant-in-Aid for Scientific Research (S) (Grant No. JP19H05600) from JSPS KAKENHI, Japan.

\end{acknowledgement}


\bibliography{opticalNiO}

\end{document}